\documentclass[pre,preprint,floatfix]{revtex4}
\usepackage{amssymb,mathptm,multirow}
\usepackage{amsmath,amssymb,amsfonts,bm}
\usepackage{graphicx,epsf,epsfig,color}
\usepackage{epsf,amsmath,amssymb,verbatim,color}
 \newcommand{\be}{\begin{equation}}
 \newcommand{\ee}{\end{equation}}
 \newcommand{\bea}{\begin{eqnarray}}
  \newcommand{\eea}{\end{eqnarray}}
  \newcommand{\im}{{\rm Im}}
   \newcommand{\re}{{\rm Re}}
\newcommand{\la}{\lambda}
\begin{document}

\title{Spectral densities of scale-free networks}

\author{D. Kim and B. Kahng}

\address{Department of Physics and Astronomy,
Seoul National University, Seoul 151-747, Korea}
\begin{abstract}
The spectral densities of the weighted Laplacian, random walk and weighted
adjacency matrices associated with a random complex network are studied using the
replica method. The link weights are parametrized by a weight exponent $\beta$.
Explicit results are obtained for scale-free networks
in the limit of large mean degree after the thermodynamic limit,
for arbitrary degree exponent and $\beta$.
\end{abstract}
\maketitle

{\bf In a network representation of complex systems, their
constituent elements and interactions between them are
represented by nodes and links of a graph, respectively.
Dynamical and structural properties of such systems can be
understood first by studying linear problems defined on the
network. A linear problem on a graph is associated with a matrix
and the distribution of its eigenvalue spectrum is of interest.
The real world networks are usually modeled as a random graph.
The spectral density, also called the density of states, is the
density of eigenvalues  averaged over an
appropriate ensemble of graph.

In this work, we study the spectral densities of several types of
matrices associated with a scale-free network which has a
power-law tail in the distribution of the number of incoming
links to a node. The spectral densities in the thermodynamic
limit are expressed in terms of solutions of corresponding
non-linear functional equations and are solved analytically in
the limit where the average incoming links per node is large.
Implications of our results are discussed.}

\section{Introduction}

Many real world networks can be modeled as a scale-free network
\cite{BarabasiA99,BoccalettiLMCH06,NewmanBW06}. In the scale-free
network, the degree $d$, the number of incident links to a node, is
distributed with a power-law tail decaying as $\sim d^{-\la}$ with
the degree exponent $\la$ often in the range $2<\la<3$.
Given such a network, one can consider several types of matrices
associated with linear problems on the network. Many structural and dynamic
properties of the network are then encoded in the eigenvalue spectra of such
matrices and hence the distributions of
their eigenvalue spectra are of interest. Since each of the real world
networks may be viewed as a realization of certain random processes,
the spectral density or the density of states is studied theoretically by
averaging them over an appropriate ensemble.

One of such an ensemble is the static model {\cite{GohKK01,LeeGKK04}
which was motivated by its simulational simplicity. Being
uncorrelated in links, it allows easier analytical treatments than
other growing type models. Other closely related one is that of
Chung and Lu \cite{ChungL02}. Recently in \cite{RodgersAKK05}, the
replica method is applied to study the spectral density of the
adjacency matrix of scale-free networks using the static model. The
expression for the spectral density is derived in terms of a
solution of a non-linear functional equation which were solved in
the dense graph limit $p\rightarrow\infty$, $p$ being the mean
degree. The explicit solution shows that the spectral density decays
as a power law with the decay exponent $\sigma_A=2\la-1$ confirming
previous approximate derivations \cite{DorogovtsevGMM03} and a
rigorous result on the Chung-Lu model \cite{ChungLV03}.

In this paper, we extend \cite{RodgersAKK05} and study three other
types of random matrices motivated from linear problems on networks.
They are the weighted Laplacian $W$, the random walk matrix $R$ and the
weighted adjacency matrix $B$, respectively. We set up the
non-linear functional equations for each type of matrices and solve
them in the dense graph limit $p\rightarrow\infty$. For the random
walk matrix, we find its spectral density to follow the semi-circle
law for all $\lambda$. For the weighted matrices, to be specific,
the weights of a link between nodes $i$ and $j$ are  given in the
form of $ (\langle d_i \rangle \langle d_j \rangle)^{-\beta/2}$
where $\langle d_i \rangle$ is the mean degree of node $i$ over the
ensemble. This form of weights is motivated by recent works on
complex networks
\cite{BarratBPV04,MacdonaldAB05,MotterZK05,ZhouMK06,Korniss06,GucluKT07,OhLKK07}.
When $\beta<1$, we find that the effect of $\beta$ on the spectral
density is to renormalize $\la$ to
$\tilde{\la}=(\la-\beta)/(1-\beta)$. The spectral density decays
with a power law with exponents $\sigma_W=\tilde{\la}$ and
$\sigma_B=(2\tilde{\la}-1)$, for $W$ and $B$, respectively for all
$\tilde{\la}>2$. When $\beta=1$, the spectral density of $B$ reduces
to the semi-circle law, the same as in $R$ while that of $W$ is bell-shaped.
When $\beta>1$, we find
that the spectral densities of $W$ and $B$ show a power-law type
singular behavior near zero eigenvalue characterized by the spectral
dimension $(\la-1)/(\beta-1)$.

This paper is organized as follows. In section 2, we generalize
\cite{RodgersAKK05} in a form applicable to other types of matrices
and present general expressions for the spectral density function in
terms of the solution of non-linear functional equation. In sections
3, 4, and 5, we define and solve the weighted Laplacian, the random
walk matrix and the weighted adjacency matrix, respectively, in the
large $p$ limit. In section 6, we summarize and discuss our results.

\section{General Formalism}

We consider an ensemble of simple graphs with $N$ nodes
characterized by the adjacency matrix $A$ whose elements
$A_{ij}=A_{ji}$ ($i\ne j$) are independently distributed with
probability \be P(A_{ij})=f_{ij} \delta(A_{ij}-1)+(1-f_{ij})
\delta(A_{ij}) \ee and $A_{ii}=0$. The degree of a node $i$ is
$d_i=\sum_j A_{ij}$ and $\langle \dots \rangle$ below denotes an
average over the ensemble.

In the static model of scale-free network \cite{GohKK01}, $f_{ij}$
is given as \be f_{ij}=1-\exp (-pNP_i P_j), \ee where $P_i \propto
i^{-1/(\la-1)}$ ($\la>2$) is the normalized weight of a node
$i=1,\ldots,N$, related to the expected degree sequence as $\langle
d_i \rangle=pNP_i$, and $p=\sum_i \langle d_i \rangle/N$ is the mean
degree of the network. The degree distribution follows the power law
$ \sim d^{-\la}$. The Erd\H{o}s-R\'{e}nyi's (ER's) classical random
graph \cite{ErdosR60} is recovered in the limit
$\la\rightarrow\infty$, where $P_i=1/N$ which is called the ER limit
below. In the model of Chung and Lu \cite{ChungL02}, $f_{ij}$ is
taken as $f_{ij}=pNP_i P_j$ , with $P_i \propto
(i+i_0)^{-1/(\la-1)}$. When $2 < \la < 3$, $i_0$ should be
$O(N^{(3-\la)/2})$ to satisfy $f_{ij}<1$ introducing an artificial
cut-off in the maximum degree. In the following, we use the static
model for ensemble averages but final results are the same for the
two models in the thermodynamic limit $N\rightarrow \infty$.

Given a real symmetric matrix $Q$ of size $N$ associated with a
graph, its spectral density, or the density of states, $\rho_Q
(\mu)$ is obtained from the formula \be \rho_Q (\mu) =
\frac{2}{N\pi} {\rm Im} \frac{\partial \langle \log Z_Q(\mu)
\rangle}{\partial \mu} ,\ee where \be
Z_Q(\mu)=\int_{-\infty}^{\infty} \left( \prod_i {\rm d} \phi_i
\right)\exp \left( \frac{{\rm i}}{2} \mu \sum_i \phi_i^2-\frac{\rm
i}{2}\sum_{ij} \phi_i Q_{ij} \phi_j \right) \label{zq1}\ee with
${\rm Im} \mu \rightarrow 0^+$. For a class of matrices considered
in this work, $Z_Q$ can be written in the form \be
Z_Q(\mu)=\int_{-\infty}^\infty \left( \prod_i {\rm d} \phi_i \right)
\exp \left(  \sum_i h_i(\phi_i) + \sum_{i<j} A_{ij} V(\phi_i,
\phi_j)\right). \label{zq} \ee

Then following \cite{KimRKK05} and \cite{RodgersAKK05} we arrive at
the expression \be \rho_Q(\mu)= \frac{2}{n \pi} {\rm
Im}\frac{\partial}{\partial \mu} \frac{1}{N} \sum_{i=1}^N \ln \int
{\rm d}^n\phi \exp \left( \sum_{\alpha} h_i(\phi_\alpha)+pNP_i \,
g_Q\{\phi_\alpha\} \right),\label{rhoq} \ee where $\alpha$=$1,
\ldots, n$ is the replica index, the limit $n \rightarrow 0$ is to
be taken after, and $g_Q\{\phi_\alpha\}$ is a solution of the
non-linear functional integral equation \be
g_Q\{\phi_\alpha\}=\sum_i P_i \frac{\int {\rm d}^n\psi  \left( \exp
(\sum_\alpha V(\phi_\alpha, \psi_\alpha)) -1 \right)\exp \left(
\sum_{\alpha} h_i(\psi_\alpha) +pNP_i \, g_Q\{\psi_\alpha\}
\right)}{\int {\rm d}^n\psi \exp \left( \sum_{\alpha}
h_i(\psi_\alpha)+pNP_i \, g_Q\{\psi_\alpha\} \right)}. \label{gq}
\ee The derivation is valid when $S_{ij}=\exp (\sum_\alpha
V(\phi_{i\alpha},\phi_{j\alpha}))-1$ satisfies the factorization
property, that is, when $S_{ij}$ can be expanded into the form \be
S_{ij}=\sum_J a_J
O_J(\phi_{i1},\ldots\phi_{in})O_J(\phi_{j1},\ldots\phi_{jn}),
\label{sij} \ee where $J$ denotes a term of the expansion, $a_J$ and
$O_J$ its coefficient and corresponding function, respectively. A
crucial step in this derivation is the use of $\log (1+f_{ij}
S_{ij}) \approx pNP_iP_j S_{ij}$. This introduces a relative error
of $\le O(N^{2-\la}\log N)$ for $2 < \la < 3$ in both the static
model and the Chung-Lu model and is neglected in the thermodynamic
limit \cite{KimRKK05,Kim07}.

If $V(\phi, \psi)$ has the rotational invariance in the replica
space, we may look for the solution of $g_Q\{\phi_\alpha\}$ in
the form of $g_Q(x)$ with $x=\sqrt{\sum_\alpha \phi_\alpha^2}$.
Then the angular integral can be evaluated and the $n \rightarrow
0$ limit can be taken explicitly. The sums over nodes are
converted to integrals using \be \frac{1}{N}\sum_i F(NP_i) =
(\la-1)\int_0^1 u^{\la-2} F
\left(\frac{(\la-2)}{(\la-1)}\frac{1}{u}\right) {\rm d} u . \ee In
the following sections, we apply this formalism to obtain formal
expressions for the spectral densities of several types of
matrices and evaluate them explicitly in the large $p$ limit.
When $\mu$ is scaled to another variable $E$, we use the
convention $\rho_Q(E)=\rho_Q(\mu)({\rm d} \mu/{\rm d} E)$ so that
$\int \rho_Q(E){\rm d} E=1$.

\section{Weighted Laplacian}

The weighted Laplacian $W$ considered in this section is
defined as \be W_{ij}= \frac{d_i \delta_{ij}-A_{ij}}{\sqrt{q_i q_j}}
\label{wij} \ee where $A$ is the adjacency matrix and $d_i =\sum_j
A_{ij}$ is the degree of node $i$, and $q_i$ are arbitrary positive
constants. This is motivated by the linear problem of the type \be
\frac{{\rm d} \phi_i}{{\rm d} t} = - \frac{1}{ q_i} \sum_j
A_{ij}(\phi_i-\phi_j) =- \sum_j \overline{W}_{ij} \phi_j,  \ee where
$\overline{W}_{ij}=(d_i \delta_{ij}-A_{ij})/q_i$. For example, in
the context of the synchronization, the input signal to a node from
its neighbors may be scaled by a factor
$d_i^{\beta}$~\cite{MotterZK05,OhLKK07}, which may be approximated
as $\langle d_i \rangle^{\beta}$~\cite{LeeJS06} or to an average
intensity of weighted networks~\cite{ZhouMK06}. Also the problem has
relevance to the Edward-Wilkinson process on network
\cite{GucluKT07}. $\overline{W}$ and $W$ are similar to each other
since $W=S^{1/2}\overline{W}S^{-1/2}$ with $S$ the diagonal matrix
with elements $S_{ii}=q_i$. Eigenvalues of $W$ are positive real
with minimum at the trivial eigenvalue 0.

When $q_i=1$ for all $i$, $W$ reduces to the standard Laplacian
$L$ defined by \be L_{ij}=d_i \delta_{ij}-A_{ij}. \ee In the
literature, the Laplacian is sometimes defined by the normalized
form \be \overline{L}_{ij}= \left\{ \begin{array}{ll}
1 & \mbox{if $i=j$ and $d_i \ne 0$}, \\
-\frac{1}{\sqrt{d_i d_j}} & \mbox{if $A_{ij}=1$}, \\
0 & \mbox{otherwise}.  \end{array} \right.  \ee We call $R \equiv
I-\overline{L}$ the random walk matrix in this work and discuss it
in the next section.
We mention here that $W$ is a weighted version of the Laplacian
of unweighted graphs while the Laplacian of weighted graphs would
have been defined by $C_{ij}=(\sum_k A_{ik}/\sqrt{q_iq_k})
\delta_{ij}  - A_{ij}/\sqrt{q_iq_j} $ \cite{ZhouMK06,Korniss06}.

For $Q=W$ in (\ref{zq1}), $Z_W(\mu)$ can be brought into the form
(\ref{zq}) by a change of variable $\phi\rightarrow\sqrt{q_i}\phi$,
with $h_i(\phi)=\frac{{\rm i}}{2} \mu q_i \phi^2$ and $V(\phi,
\psi)=-\frac{{\rm i}}{2}(\phi -\psi)^2$. Inserting these into
(\ref{rhoq}) and (\ref{gq}), and evaluating the angular integral, we
obtain \be \rho_W(\mu)=\frac{1}{\pi N} \re \sum_i q_i \int_0^\infty
y \exp(\frac{{\rm i}}{2} \mu  q_i y^2 + pNP_i \, g_W(y))  {\rm d} y
\label{rhow0} \ee with
\bea g_W (x) &=& {\rm e}^{-{\rm i} x^2 /2}-1 \nonumber \\
& & -x {\rm e}^{-{\rm i} x^2 /2}\sum_i P_i \int_0^\infty   J_1(xy)
\exp(\frac{\rm i}{2} \mu q_i y^2 -\frac{\rm i}{2} y^2
  + pNP_i \, g_W(y))  {\rm d} y, \label{gw} \eea
where $J_1(z)$ is the Bessel function of order one. In the ER limit
($NP_i=1$) and $q_i=1$, (\ref{rhow0}) and (\ref{gw}) reduce to
equations (17) and (16) of \cite{BrayR88}, respectively.

The dense graph limit $p\rightarrow\infty$ is investigated by using
the scaled function $G_W(x)=pg_W(x/\sqrt{p})$. Then in the limit
$p\rightarrow\infty$, $G_W(x)=-\frac{\rm i}{2}x^2$ and \be
\rho_W(\mu)=\frac{1}{\pi N} \im \sum_i \frac{q_i}{pNP_i-\mu q_i} \ee
for arbitrary $q_i$. To be specific, we now set $q_i=\langle d_i
\rangle^\beta=(pNP_i)^\beta$ at this stage. $\beta$ is arbitrary and
is called the weight exponent here. When the eigenvalue is scaled as
\be E=p^{\beta-1} (\la-2)^{\beta-1}(\la-1)^{1-\beta} \mu, \ee we
find the spectral density in the dense graph limit as \be
\rho_W(E)=\frac{\la-1}{\pi} \im
\int_0^1\frac{u^{\la-\beta-1}}{1-u^{1-\beta}E} {\rm d} u.
\label{rhow}\ee This shows qualitatively different behaviors in the
three regions of $\beta$.

\begin{figure}
\epsfig{file=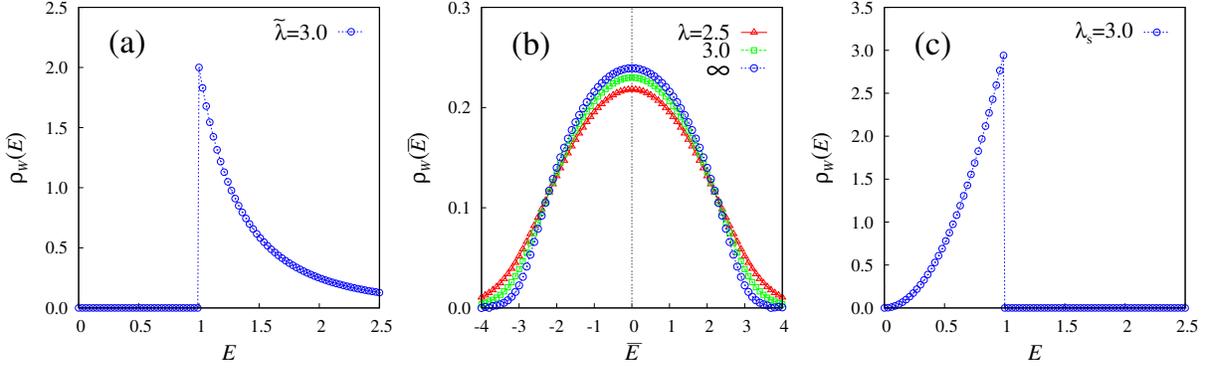,width=5cm,angle=270} \caption{(Color online)
The spectral density of the weighted Laplacian
for weight exponent $\beta <1$ (a), $\beta=1$ (b), and $\beta > 1$ (c).
In (a) and (c), a typical curve is shown as a function of $E$ while in (b),
the spectral density is shown as a function of $\overline{E}=\sqrt{p}(E-1)$
for several values of $\la$. }
 \label{fig1}
\end{figure}

(i) $\beta<1$:

Eq.~(\ref{rhow}) is evaluated as \be \rho_W(E)=\left\{
\begin{array}{ll}
0 & \mbox{if $0<E<1$}, \\(\frac{\la-1}{1-\beta})
E^{-(\la-\beta)/(1-\beta)} & \mbox{if $E>1$}. \label{rhow1}
\end{array} \right. \ee
Therefore, the spectra has a finite gap in $E$ and a power-law tail
with an exponent $\sigma_W=(\la-\beta)/(1-\beta)$. The only effect
of $\beta$ here is to renormalize $\la$ to \be
\tilde{\la}=\frac{\la-\beta}{1-\beta}. \label{latilde} \ee
Fig.~\ref{fig1}(a) shows the graph of $\rho_W(E)$ for
$\tilde{\la}=3$.

(ii) $\beta=1$:

The case $\beta=1$ needs a special treatment.
When $\beta=1$, $E=\mu$ and $\rho_W(E)=\delta(E-1)$. However, if
we expand the region near $E=1$  by introducing a new variable
$\overline{E}$ by $\overline{E}=\sqrt{p}(E-1)$, we obtain
non-trivial values for finite $\overline{E}$. The method and result are
similar to that treated in~\cite{BrayR88} for the ER case.
Following \cite{BrayR88}, the function $g_W(x)$ in Eq. (15) is
written in terms of $H(x)$, defined by \be pg_W(p^{-1/4}x)=-{\rm
i}\sqrt{p}x^2/2-x^4/8+H(x).\ee  Then in the limit
$p\rightarrow\infty$, (\ref{gw}) gives $H(x)={\rm i} \gamma(\overline{E})
 x^2/2$ and (\ref{rhow0}) gives $\rho_W(\overline{E})=\im
\gamma(\overline{E})/\pi$, where $\gamma(\overline{E})$ is the solution
of \be \gamma=-\frac{(\la-1)}{\sqrt{2\pi}}\int_{-\infty}^\infty {\rm
d} z {\rm e}^{-z^2/2}\int_0^1 \frac{u^{\la-2} {\rm d}
u}{\overline{E}+\gamma+z\sqrt{u(\la-1)/(\la-2)}} .\ee
Fig.~\ref{fig1}(b) shows the graph of $\rho_W(\overline{E})$ for several
values of $\lambda$.

(iii) $\beta>1$:

In this case, (\ref{rhow}) is evaluated as \be \rho_W(E)=\left\{
\begin{array}{ll} (\frac{\la-1}{\beta-1}) E^{(\la-\beta)/(\beta-1)} &
\mbox{if $0<E<1$} \\ 0 & \mbox{if $E>1$}.
\end{array} \right. \label{rhow2}\ee
Thus $\rho_W(E)$ is non-zero for $0<E<1$ with a simple power. Such
power-law dependence near the zero eigenvalue gives arise a
long-time relaxation $\sim t^{-\lambda_s}$ with the spectral
dimension~\cite{RammalT83,AlexanderO82,BurdaCK01,DestriD02} \be
\la_s=\frac{\la-1}{\beta-1}. \label{las} \ee Fig.\ref{fig1}(c) shows
the graph of $\rho_W(E)$ for $\la_s=3$.

\section{Random walk matrix}

Consider a random walk problem defined as follows: When a random
walker at a node $i$ sees $d_i$ neighbors, it jumps to one of them
with equal probability. When a node $i$ is isolated so that $d_i=0$,
the random walker is supposed not to move. Then the transition
probability from node $i$ to $j$ is given by $\overline{R}_{ij} =
A_{ij}/d_i$ if $d_i \ne 0$ and $\overline{R}_{ij}=\delta_{ij}$ if
$d_i=0$. $\overline{R}$ can be brought into a symmetric form by a
similarity transformation $R=T^{1/2} \overline{R} T^{-1/2}$, where
$T$ is the diagonal matrix with elements $T_{ii}=1$ when $d_i=0$ and
$T_{ii}=d_i$ when $d_i\ne0$. The resulting symmetric matrix $R$,
called the random walk matrix here, is \be R_{ij}= \left\{
\begin{array}{ll}
1  & \mbox{if $i=j$ and $d_i =0$}, \\
\frac{1}{\sqrt{d_i d_j} } & \mbox{if $A_{ij}=1$}, \\
0 & \mbox{otherwise.}  \end{array} \right. \label{rij} \ee Being
similar, $R$ and $\overline{R}$ have the same set of eigenvalues
that are located within the range $|\mu|\le1$.

The isolated nodes integrate out of the partition function
$Z_R(\mu)$ in (\ref{zq1}), giving an additive term $n_0
\delta(\mu-1)$ in the spectral density $\rho_R(\mu)$ where $n_0$ is
the density of isolated nodes. For the remaining nodes, $R_{ij}=
A_{ij}/\sqrt{d_i d_j}$ and, after a change of variable $\phi_i
\rightarrow \sqrt{d_i} \phi_i$, $Z_R(\mu)$ can be brought into the
form (\ref{zq}) with $V(\phi, \psi)=\frac{\rm i}{2} \mu
(\phi^2+\psi^2)-{\rm i} \phi\psi$ and $h_i(\phi)=-\epsilon \phi^2 $
with $\epsilon \rightarrow 0^+$ to ensure convergences. Since
$A_{ij}=0$ anyway for isolated nodes, the sums in (\ref{zq}) is
extended to all nodes. Plugging this into (\ref{rhoq}) and
(\ref{gq}), we find \be \rho_R (\mu)= n_0
\delta(\mu-1)+\frac{p}{\pi} \re \sum_i P_i \int_0^\infty y g_R(y)
\exp (p N P_i [g_R(y)-1]) {\rm d} y \label{rhor} \ee with \be
g_R(x)={\rm e}^{{\rm i} \mu x^2 /2}-x \sum_i P_i \int_0^\infty J_1
(xy) \exp ( \frac{\rm i}{2}\mu (x^2+y^2)+pNP_i [g_R(y)-1])  {\rm d}
y \label{gr}.\ee

When $p\rightarrow\infty$, all nodes belong to the percolating giant
cluster \cite{LeeGKK04} and $n_0$ vanishes. To obtain the spectral
density in the limit $p\rightarrow\infty$, we scale $\mu=p^{-1/2} E$
and $g_R(x)=1+G_R(p^{1/4}x)/p$. Then from (\ref{gr}), $G_R(x)$ is
determined as $G_R(x)=-a x^2/2$ with $a$ being a solution of
$a^2+{\rm i} E a -1=0$ and from (\ref{rhor}), $\rho_R(E)=\re (\pi
a)^{-1}$. This gives the semi-circle law for all $\la$: \be
\rho_R(E) = \frac{1}{\pi} \sqrt{1-\frac{E^2}{4}} \ee for $|E|\le2$
and 0 otherwise.

\section{Weighted adjacency matrix}

In this section, we consider the weighted version of $A$ defined by
\be B_{ij}=\frac{A_{ij}}{\sqrt{q_i q_j}},  \label{bij} \ee where
$q_i$ are arbitrary positive constants. This is motivated by the
weighted networks whose link weights are product of quantities
associated with the two nodes at each end of the
link\cite{BarratBPV04,MacdonaldAB05}. Later on for explicit
evaluations, we take $q_i$ to be $q_i = \langle d_i \rangle^\beta =
(pNP_i)^\beta$ with arbitrary $\beta$. When $\beta=0$, we recover
$A$ treated in \cite{RodgersAKK05} while when $\beta=1$,
$B_{ij}=A_{ij}/\sqrt{\langle d_i \rangle \langle d_j \rangle}$ may
be considered as an approximation to $R$ and is treated in
\cite{ChungLV03}. With a change of variable $\phi_i \rightarrow
\sqrt{q_i}\phi_i$ in (\ref{zq1}), $Z_B$ is of the form (\ref{zq})
with $h_i(\phi)=\frac{\rm i}{2} \mu q_i \phi^2$ and
$V(\phi,\psi)=-{\rm i} \phi \psi$. Then we find \be \rho_B
(\mu)=\frac{1}{\pi N} \re \sum_i q_i \int_0^\infty y \exp(\frac{\rm
i}{2} \mu q_i y^2+pNP_i \, g_B(y)) {\rm d} y  \label{rhob} \ee with
\be g_B (x) = - \sum_i P_i \int_0^\infty   x J_1(xy) \exp(\frac{\rm
i}{2} \mu q_i y^2
  + pNP_i g_B(y)) {\rm d} y \label{gb} \ee
for arbitrary $q_i$.

Specializing to the case where $q_i=\langle d_i
\rangle^\beta=(pNP_i)^\beta$, the limit $p\rightarrow\infty$ is
investigated by scaling $\mu$ to $E$ by \be
E=p^{\beta-\frac{1}{2}}(\la-2)^{\beta-1}(\la-1)^{1-\beta}\mu,\ee and
$g_B(x)=G_B(p^{1/4}x)/p$. Then $G_B(x)= -\frac{\rm i}{2} E b(E) x^2$
and \be \rho_B(E)=-\frac{E}{\pi} \im b^2(E) \ee with  $b(E)$ as the
solution of \be E^2 b = (\la-1)\int_0^1
\frac{u^{\la-2}}{u^{1-\beta}-b} {\rm d} u. \label{bofb} \ee In the
ER limit, we recover the semi-circle law regardless of $\beta$. For
finite $\la$, we consider three regions of $\beta$ separately.

(i) $ \beta<1$:

A change of integration variable in (\ref{bofb}) leads to \be \
_2F_1\left(1, \frac{\la-1}{1-\beta};\frac{\la-\beta}{1-\beta};
\frac{1}{b}\right)=-b^2 E^2  \label{bofb1} \ee where $\
_2F_1(1,c-1;c;z)=(c-1)\int_0^1t^{c-2}/(1-tz){\rm d} z$ is the
hypergeometric function. Eq. (\ref{bofb1}) is a generalization of
\cite{RodgersAKK05} which is a special case of $\beta=0$. One notes
that the effect of $\beta$ is again to renormalize $\la$ to $
\tilde{\la} =(\la-\beta)/(1-\beta)$ and the results of
\cite{RodgersAKK05} applies here when its degree exponent is
replaced by the effective one. In particular, the spectral density
is symmetric in $E$, has the power-law tail $\sim |E|^{-\sigma_B}$
 with an exponent
\be \sigma_B = 2\tilde{\la}-1= \frac{2\la-\beta-1}{1-\beta}  \ee and
an analytic maximum $(\tilde{\la}-1)/(\tilde{\la}\pi)$ at $E=0$.
Fig.~\ref{fig2}(a) shows the graph of $\rho_B(E)$ for several values
of $\tilde{\la}$.

\begin{figure}
\epsfig{file=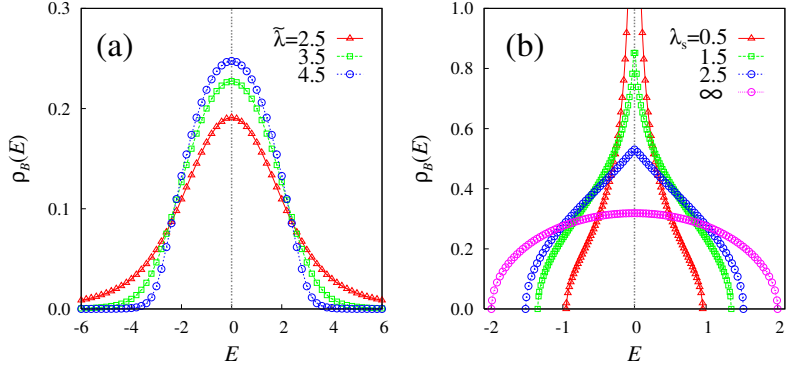,width=5cm,angle=270} \caption{(Color online)
The spectral density of the weighted adjacency matrix with weight
exponent $\beta<1$ in (a) and $\beta > 1$ in (b), for several values
of effective degree exponents $\tilde{\la}$ (a) and $\lambda_s$ (b),
respectively.}\label{fig2}
\end{figure}

(ii) $\beta=1$:

In this case, $b$ in (\ref{bofb}) is simply determined from $E^2
b=1/(1-b)$ and the spectral density becomes the semi-circle law:
\be \rho_B(E) = \frac{1}{\pi} \sqrt{1-\frac{E^2}{4}} \ee for
$|E|\le 2$ and 0 for $|E|>2$ for all $\la$. The same is proved in
~\cite{ChungLV03} for sufficiently large but finite $p$ while we
have taken the limit $p\rightarrow\infty$. $B$ at $\beta=1$ being
an approximation of $R$, it is not surprising to see the same
results for the two cases.

(iii) $\beta>1$:

In this case, it is convenient to bring (\ref{bofb}) into  the form
\be E^2 =\left( \frac{\la_s}{\la_s+1}\right) \frac{1}{b} \ _2F_1
(1,\la_s+1;\la_s+2;b) . \label{bofb2} \ee with $
\la_s=(\la-1)/(\beta-1)$. The right hand-side of (\ref{bofb2}) as a
function of real $b$ takes a minimum value $E_c^2$ at $0<b_c<1$ and
increases to infinity as $b\rightarrow0^+$ or $b\rightarrow1^-$.
Thus for $|E|>E_c$, $b(E)$ is real and $\rho_B(E)=0$. As $|E|$
decreases from $E_c$, $\rho_B(E)$ rises with a square root
singularity since the righthand side of (\ref{bofb2}) is analytic at
$b_c$. It is interesting to note that the behavior of $\rho_B(E)$ at
$E=0$ is non-analytic. When $0<\la_s<1$, it diverges as \be
\rho_B(E) \sim \frac{\la_s}{2 \cos (\frac{\pi}{2}\la_s)}
|E|^{-(1-\la_s)} \ee while, for $\la_s>1$, its singular part is
masked by the analytic part and it takes the finite maximum value
\be \rho_B(0)= \frac{1}{\pi} \frac{\la_s}{\la_s-1} . \ee At
$\la_s=1$, it diverges logarithmically: \be  \rho_B(E) \sim
\frac{1}{\pi} \log \frac{1}{|E|}. \ee Fig.~\ref{fig2}(b) shows the
graph of $\rho_B(E)$ for several values of $\la_s$.

\section{Summary and Discussion}

In this work, we derived the spectral densities of three types of
random matrices, the weighted Laplacian $W$, the random walk
matrix $R$, and the weighted adjacency matrix $B$, of the static
model in the dense graph limit after the thermodynamic limit. Our
results apply to the model of Chung-Lu also. In fact, they apply
to other models as long as $f_{ij}$ in (1) is a function of $pNP_iP_j$ and
satisfy $f_{ij}\le pNP_iP_j$.

With weights of the form $q_i=\langle d_i \rangle^{\beta}$, they
show varying behaviors depending on the degree exponent $\la$ and
the weight exponent $\beta$. The spectrum follows the semi-circle
law for $R$, and at the $\beta=1$ point of $B$ for all $\lambda$.
The $\beta=1$ point of $W$ is closely related to $I-B$ or $I-R$, but its
spectral density is not of the semi-circle law but is bell-shaped.
When $\beta<1$, the degree exponent is renormalized to
$\tilde{\la}$ given in (\ref{latilde}) and the spectral density
shows a power law decay with exponent $\sigma_W=\tilde{\la}$ and
$\sigma_B=2\tilde{\la}-1$  for $W$ and $B$, respectively. When the
eigenvalue spectrum has a long tail decaying as $\sim
\mu^{-\sigma}$, the maximum eigenvalue of a finite system is
expected to scale with $N$ as $\mu_{\rm N} \sim N^{1/(\sigma-1)}$
while the natural cutoff of degree in the scale-free network is
$d_{\rm max} \sim N^{1/(\la-1)}$. The maximum eigenvalue of the
weighted Laplacian $W$ may be taken as $W_{11} \sim d_{\rm
max}^{1-\beta} $ for $\beta<1$ in the first order perturbation
approximation.  This simple argument explains the power of the
tail $\sigma_W=\tilde{\la}=(\la-\beta)/(1-\beta)$ for $W$. A
similar argument applied to $(B^2)_{11}\sim d_{\rm
max}^{1-\beta}$ gives the decay exponent
$\sigma_B=2\tilde{\la}-1$ for $B$.

When $\beta>1$, the spectral densities of $W$ and $B$ are non-zero
within a finite interval of the scaled eigenvalue $E$ and are
associated with the spectral dimension $\la_s$ given in (\ref{las}).
For $W$, it is a simply power $\sim E^{\la_s-1}$ in $0<E<1$, while
for $B$, it is symmetric in $E$ and singular at $|E|=0$ with
exponent $\la_s-1$. They both diverge as $E\rightarrow0$ when
$0<\la_s<1$.

When $p$ is finite, the spectra is very complicated and is not well
understood. For small $p$ at least, one expects infinite number of
delta peaks on the spectrum \cite{BauerG01}. In the dense graph
limit $p\rightarrow\infty$, those delta peaks have disappeared. Even
though our explicit results are for the limit $p\rightarrow\infty$,
the limit is taken after the thermodynamics limit
$N\rightarrow\infty$, and physically they would be a good
approximation for $1 \ll p \ll N$ in finite systems. In the
synchronization problem on networks, the eigenratio $R=\mu_N/\mu_2$
of $W$ is of interest~\cite{BarahonaP02}. From (\ref{rhow1}), one
may estimate $\mu_N\sim N^{{(1-\beta)}/{(\lambda-1)}}$ and $\ln R
\sim \frac{1-\beta}{\lambda-1}\ln N$ for $\beta < 1$, assuming that
$N$-dependence of $\mu_2$ is slower than the power law. Similarly,
from (\ref{rhow2}), one gets $\ln R \sim
\frac{\beta-1}{\lambda-1}\ln N$ for $\beta > 1$. Such
$\beta$-dependence of $R$ is corroborated with numerical results for
a similar matrix studied in~\cite{MotterZK05}.

The spectral properties of Laplacian on weighted networks,
$C_{ij}=(\sum_k B_{ik})\delta_{ij}-B_{ij}$ or its normalized version
$ D_{ij}=\delta_{ij}-B_{ij}/\sqrt{\sum_k B_{ik} \sum_k B_{jk}} $ are
also of interest\cite{ZhouMK06,Korniss06}. Unfortunately, the
formalism leading to (\ref{rhoq}) and (\ref{gq}) cannot be applied
to these cases since the factorization property (\ref{sij}) is not
satisfied.

\section*{Acknowledgement}
The authors thank G. J. Rodgers, Z. Burda and J. D. Noh for useful
comments. This work is supported by KRF Grant No.
R14-2002-059-010000-0 of the ABRL program funded by the Korean
government (MOEHRD).

\end{document}